\documentclass[a4paper,12pt]{article} \textheight 23cm
\usepackage{graphicx}
\usepackage{amsmath}

\usepackage{color}

\begin{document}

\newpage
%opening
\title{Relation between Lacunarity, Correlation dimension and Euclidean dimension of Systems.}
\author{Abhra Giri$^{1,2}$, Sujata Tarafdar$^2$, Tapati Dutta$^{1,2*}$}
\maketitle
\noindent
$^1$Physics Department, St. Xavier's College, Kolkata 700016, India\\
$^2$Condensed Matter Physics Research Centre, Physics Department, Jadavpur University, Kolkata 700032, India\\
\noindent
${^*}$ Corresponding author: Email: tapati$\_$mithu@yahoo.com\\ Phone:+919330802208, Fax No. 91-033-2287-9966

\newpage
\noindent{\bf Abstract} 
Lacunarity is a measure often used to quantify the lack of translational invariance present in fractals and multifractal systems. The generalised dimensions, specially the first three, are also often used to describe various aspects of mass distribution in such systems. In this work we establish that the graph (\textit{lacunarity curve}) depicting the variation of lacunarity with scaling size, is non-linear in multifractal systems. We propose a generalised relation between the Euclidean dimension, the Correlation Dimension and the lacunarity of a system that lacks translational invariance, through the slope of the lacunarity curve. Starting from the basic definitions of these measures and using statistical mechanics, we track the standard algorithms- the box counting algorithm for the determination of the generalised dimensions, and the gliding box algorithm for lacunarity, to establish this relation.  We go on to validate the relation on six systems, three of which are deterministically determined, while three others are real. Our examples span 2- and 3- dimensional systems, and euclidean, monofractal and multifractal geometries. We first determine the lacunarity of these systems using the gliding box algorithm. In each of the six cases studied, the euclidean dimension, the correlation dimension in case of multifractals, and the lacunarity of the system, together, yield a value of the slope $S$ of the lacunarity curve at any length scale. The predicted $S$ value matches the slope as determined from the actual plot of the lacunarity curves at the corresponding length scales. This establishes that the relation  holds for systems of any geometry or dimension.

\vskip 0.5cm

\section{Introduction}
Spatial patterns observed in various branches of natural sciences as in geology, biology and ecology are often complex as they exhibit scale-dependent structures. When the systems show self-similarity, their structure is described by the fractal dimension $D_0$\cite{mandelbrot}. In reality complex structures exist with subsets of region having different scaling properties and can rarely be described as perfect monofractals. In many cases multifractal analysis provides more information about the space filling properties than the fractal dimension $D_0$ alone. Moreover there are several systems that have the same fractal dimension but have different texture. Lacunarity is a parameter that quantifies this difference in the texture of materials having the same fractal dimension. It describes the distribution of the sizes of gaps or lacunae
surrounding an object within a image. The lacunarity $\Lambda$, quantifies the extent to which a fractal fails to be translationally invariant. Its value decreases as the system becomes more homogeneous. While
lacunarity can be used to distinguish objects with similar fractal dimensions but whose structure fill the space differently, it can also be used independently as a general tool for describing spatial patterns. 
  The concept of lacunarity  has been used by many \cite{Allain1991, Plotnick1996, Lin1986} to describe the micro-structural pattern of systems. Gefen et al.\cite{Gefen1984} have shown that lacunarity plays an important role in critical phenomena and may be used to classify the universality class.

 The lacunarity $\Lambda$ is a function of $\epsilon/L$, 
 where $\epsilon$ is the size of the scaling box and $L$ the system size. The \textit{lacunarity curve } which plots $ln(\Lambda)$ against $ln(\epsilon/L)$ is in general non-linear. A linear lacunarity curve is obtained only for a system with monofractal scaling. In this paper we show that a multifractal distribution does not in general lead to a linear lacunarity curve  
and propose a generalised relation between the lacunarity $\Lambda$, the suitable generalised dimension $D$ and the euclidean dimension $E$ which will hold true for all systems irrespective of the geometry - euclidean, monofractal or multifractal. We have tested the validity of the relation for 2-dimensional and 3-dimensional systems and found it to hold true. 

According to Allain et al. \cite{Allain1991} the lacunarity $\Lambda$ follows a power law with $\epsilon/L$, with exponent  $(D-E)$ where $D$ and $E$ are the generalised dimension and the Euclidean dimension of the system. For a monofractal $D=D_0$, the fractal dimension, whereas for a multifractal $D=D_2$, the correlation dimension. In that case the lacunarity curve would be linear for a mono-fractal as well as a multi-fractal distribution. Our analysis leads to additional terms, not present in Allain's expression for lacunarity, which introduces non-linearity in the lacunarity curve for multi-fractal systems. 
 We quantify the lacunarity of six systems - one euclidean, one monofractal and four multifractal systems. Of the four multifractal systems studied, one is determined deterministically while the other three are real systems - sedimentary rocks of low and high porosity\cite{gji2015}, and crystal aggregates grown from solution of salt and gelatin\cite{cgd2012,colsua2013}. 
We have followed the procedure for multifractal analysis as elaborated in standard texts on fractals \cite{Vicsek, Feder, Diaz} to determine the generalized dimension of the multifractal systems whose lacunarity we calculated. The details of the procedure and results of the multifractal analysis are to be found in \cite{gji2015, cgd2012, colsua2013}. To determine the lacunarity of the systems studied, we have followed the gliding box algorithm of Allain et al.\cite{Allain1991}. In the following section we give a brief outline of the algorithm and the lacunarity analysis of systems having different geometries and dimensions.  We formulate the proposed relationship between the lacunarity $\Lambda$ and the generalised dimension of a system in section(\ref{sec2}). The proposed relation is then validated for the different systems we studied.

\section{Lacunarity using the gliding box algorithm}
In this method a square box of size $\epsilon$ glides over the system and counts the number of pixels $m$ inside the box. Let n(m,$\epsilon$) be the number of boxes of size $\epsilon$ and containing mass m.
The probability function for such a box is defined as \cite{Allain1991, Plotnick1996} 
\begin{equation}
Q(m,\epsilon)=\frac{n(m,\epsilon)}{\sum_{m=1}^{\epsilon^{2}} n(m,\epsilon)}.
\end{equation}
The first and second moments of this distribution are given by
\begin{equation}
Z(1)= \sum m Q(m,\epsilon)
\end{equation}
\begin{equation}
Z(2)= \sum m^{2} Q(m,\epsilon)
\end{equation}
The lacunarity at scale $\epsilon$ is then defined as
\begin{equation}
\Lambda(\epsilon)=\frac{\sum_{m=1}^{\epsilon^2} m^{2}Q(m,\epsilon)}{\Big[\sum_{m=1}^{\epsilon^{2}} mQ(m,\epsilon)\Big]^{2}} 
\label{lacuna}
\end{equation}
i.e.
\begin{equation}
\Lambda(\epsilon)=\frac{Z(2)}{[Z(1)]^{2}}
\label{lacuna1}
\end{equation}
The statistical interpretation of lacunarity is better understood by recalling
\begin{equation}
Z(1)= <m(\epsilon)>
\end{equation}
\begin{equation}
Z(2)= m_{v}^{2}(\epsilon) + <m^{2}(\epsilon)>
\end{equation}
where $<m(\epsilon)>$ and $m_{v}^{2}(\epsilon)$ are the mean and the variance of mass in each box.
Therefore from eq.(\ref{lacuna})
\begin{equation}
\Lambda(\epsilon)= m_{v}^{2}(\epsilon)/<m^{2}(\epsilon)> + 1
\label{lacu-interp}
\end{equation}
Thus lacunarity is a dimensionless ratio of the variance to the mean value of the mass distribution. From an inspection of eq.(\ref{lacu-interp}) it is clear that as the mean Z(1) goes to zero, lacunarity $\Lambda$ tends to $\infty$, i.e. sparse sets have higher lacunarity than denser ones. Lacunarity $\Lambda$ also depends on the box size $\epsilon$. For a system with a highly clustered mass distribution, boxes with larger $\epsilon$ will be more translationally invariant than boxes of smaller $\epsilon$, i.e. Z(2) decreases with respect to Z(1). Therefore the same mass distribution will have lower lacunarity as the box size increases.

\section{Relation between generalised dimension and  lacunarity}\label{sec2}

In this section we present our derivation of a relation connecting the lacunarity, generalised dimension and the euclidean dimension of a system. The results are slightly different from the expression obtained by Allain et al.\cite{Allain1991}

Following the box-counting algorithm, let the whole structure of size L be covered by boxes of size $\epsilon$ and the mass of $i$-th box be $\textsc{m}(i,\epsilon)$.

So the total mass is
$\textsc{M}(\epsilon)=\displaystyle\sum_{i=1}^{N(\epsilon)}\textsc{m}(i,\epsilon)$,
where $N(\epsilon)$ is the  number of boxes required to cover the whole structure.

The total mass distribution raised to power $q$ for a particular box size $\epsilon$ is given by,
\begin{equation}
I(q,\epsilon)=\displaystyle\sum_{i=1}^{N(\epsilon)}\Big[\frac{\textsc{m}(i,\epsilon)}{\textsc{M}(\epsilon)}\Big]^q 
\label{iq}
\end{equation}
where $q\in\Re$.

For different values of $\epsilon$, the slope of the plot $I(q,\epsilon)$ vs. $\epsilon/L$ in log-log scale is given by,
\begin{equation}
\tau(q)=-\frac{lnI(q,\epsilon)}{ln(\frac{\epsilon}{L})}
\label{tau}
\end{equation}

For a multifractal system, the generalised dimension for any moment $q$ is defined by \cite{Feder}
\begin{equation}
D_q=\frac{\tau(q)}{1-q}
\label{dq}
\end{equation} 
  The generalised dimensions $D_q$ for $q=0$, $q=1$ and $q=2$ are known as the Capacity, the Information (Shannon entropy) and 
Correlation Dimensions respectively  \cite{Shannon1949, Grassberger1983}.
The Correlation Dimension $D_2$ describes the uniformity of the measure (here mass) in different intervals.
Smaller $D_2$ values indicate long-range dependence, whereas higher values indicate domination of short range dependence.
From eqs.(\ref{tau} and \ref{dq}),  with $q=2$, the Correlation dimension is given by 
\begin{equation}
D_2=\frac{\tau(2)}{1-2}=-\tau(2)=\frac{lnI(2,\epsilon)}{ln(\frac{\epsilon}{L})}
\label{d2}
\end{equation}
Substituting eq.(\ref{iq}) in eq.(\ref{d2}) we get,
\begin{equation}
D_2=\frac{ln\displaystyle\sum_{i=1}^{N(\epsilon)}\Big[\frac{\textsc{m}(i,\epsilon)}{\textsc{M}(\epsilon)}\Big]^2}{ln(\frac{\epsilon}{L})} 
\end{equation}

Further, let the number of boxes containing mass $m$ and of size $\epsilon$ be $B(m,\epsilon)$. The corresponding probability is,
\begin{equation}
P(m,\epsilon)=\frac{B(m,\epsilon)}{N(\epsilon)}
\label{P} 
\end{equation}
Using the definition of lacunarity as given by eq.(\ref{lacuna}) and the above equation, the lacunarity of the structure may be written as,
\begin{equation}
\Lambda(\epsilon)=\frac{\displaystyle\sum_{m=1}^{M(\epsilon)}m^2P(m,\epsilon)}{\Big[\displaystyle\sum_{m=1}^{M(\epsilon)}mP(m,\epsilon)\Big]^2} 
\label{lacuna2}
\end{equation} 

Substituting eq.(\ref{P}) we get,
\begin{equation}
\Lambda(\epsilon)=\frac{\displaystyle\sum_{m=1}^{M(\epsilon)}m^2\frac{B(m,\epsilon)}{N(\epsilon)}}{\Big[\displaystyle\sum_{m=1}^{M(\epsilon)}m\frac{B(m,\epsilon)}{N(\epsilon)}\Big]^2} 
\label{lacuna3}
\end{equation}
 
Since the total number of boxes of size $\epsilon$  is $N(\epsilon)=(\frac{L}{\epsilon})^E$ and it does not depend on mass, eq.(\ref{lacuna3}) may be written as
\begin{equation}
\Lambda(\epsilon)=N(\epsilon)\frac{\displaystyle\sum_{m=1}^{M(\epsilon)}m^2B(m,\epsilon)}{\Big[\displaystyle\sum_{m=1}^{M(\epsilon)}mB(m,\epsilon)\Big]^2}=(\frac{L}{\epsilon})^E\frac{\displaystyle\sum_{m=1}^{M(\epsilon)}m^2B(m,\epsilon)}{\Big[\displaystyle\sum_{m=1}^{M(\epsilon)}mB(m,\epsilon)\Big]^2} 
\label{lacuna4}
\end{equation}
where $E$ is the Euclidean dimension of the structure and $L$ is the system size.

From eq.(\ref{lacuna4}), it follows that the slope of $\Lambda(\epsilon)$ vs. $\epsilon/L$ plot in a log-log  plot is given by,
%%%%%%%%%%%%%%%%%%%%%%%%%%%%%%%%%%%%%%%
\begin{equation}
\begin{aligned}
S       &= \frac{ln\Lambda(\epsilon)}{ln(\frac{\epsilon}{L})} = \frac{ln\Bigg\lbrace(\frac{L}{\epsilon})^E\frac{\displaystyle\sum_{m=1}^{M(\epsilon)}m^2B(m,\epsilon)}{\Big[\displaystyle\sum_{m=1}^{M(\epsilon)}mB(m,\epsilon)\Big]^2}\Bigg\rbrace}{ln(\frac{\epsilon}{L})}\\
        &= E\frac{ln(\frac{L}{\epsilon})}{ln(\frac{\epsilon}{L})}+\frac{ln\Bigg\lbrace\frac{\displaystyle\sum_{m=1}^{M(\epsilon)}m^2B(m,\epsilon)}{\Big[\displaystyle\sum_{m=1}^{M(\epsilon)}mB(m,\epsilon)\Big]^2}\Bigg\rbrace}{ln(\frac{\epsilon}{L})}
\end{aligned}
\label{slope}
\end{equation}
%%=====================================
 Reorganising eq.(\ref{slope}),
\begin{equation}
S=-E+\frac{ln\Bigg\lbrace\frac{\displaystyle\sum_{m=1}^{M(\epsilon)}m^2B(m,\epsilon)}{\Big[\displaystyle\sum_{m=1}^{M(\epsilon)}mB(m,\epsilon)\Big]^2}\Bigg\rbrace}{ln(\frac{\epsilon}{L})}
\label{slope1}
\end{equation}
\\

\begin{figure}[!h]
\begin{center}
\includegraphics[scale=0.4]{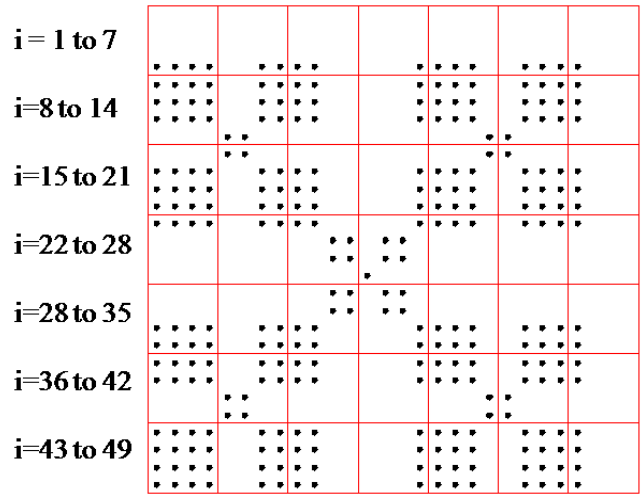} 
\caption{Index $i$ denotes box label, a single dot represents unit mass. }\label{vicsek}
\end{center}
\end{figure}
%\\
\begin{figure}[!h]
\begin{center}
\includegraphics[scale=0.4]{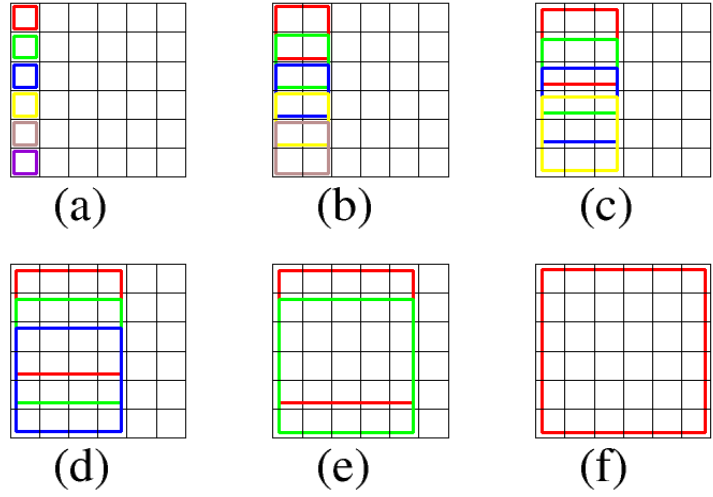} 
\caption{Figure displays that $N^\prime(\epsilon)= (L-\epsilon+1)^E$ for gliding box. The panels show this for different values of $\epsilon$. }\label{glb}
\end{center}
\end{figure}

To establish the connection between the box counting method used for multifractal analysis and the gliding box algorithm used to determine the lacunarity of the same system, let us examine fig.(\ref{vicsek}).

As an example, let us take a square box of size $\epsilon=4$ to cover the system. Each dot represents one unit of mass. Let us calculate $\displaystyle\sum_{m=1}^{M(\epsilon)}m^2B(m,\epsilon)$ of eq.(\ref{slope1}) for $m=12$.  Here $B(12,4)=4$. Hence for $m=12,  m^2B(m,\epsilon)=12^2*4$.

According to the box counting method, the boxes with indices $i=8,15,47,48$ have the mass $ m=12$. Therefore, $\displaystyle\sum_{i=1}^{N(\epsilon)}\textsc{m}(i,4)^2=4*12^2$. Examination of the fig.(\ref{vicsek}) reveals that  
$ \displaystyle\sum_{i=1}^{N(\epsilon)}\Big[\textsc{m}(i,\epsilon)\Big]^2= \displaystyle\sum_{m=1}^{M(\epsilon)}m^2B(m,\epsilon) $ for any amount of mass considered. 
  Examination of the same figure shows \\
$ \textsc{M}(\epsilon)=\displaystyle\sum_{i=1}^{N(\epsilon)}\textsc{m}(i,\epsilon)= \displaystyle\sum_{m=1}^{M(\epsilon)}mB(m,\epsilon) $.
Substituting eq.(\ref{d2} ) in  eq.(\ref{slope1}) we get,\\
\begin{equation}
S=-E+D_2
\label{slope2}
\end{equation}

If we calculate lacunarity using gliding box algorithm, the number of boxes containing mass $m$ and of size $\epsilon$, $B^\prime(m,\epsilon)$, is greater than $B(m,\epsilon)$, let us say by an amount $\textsc{B}(m,\epsilon)$.
Also the number total of boxes of size $\epsilon$ is $N^\prime(\epsilon)$, which is greater than $N(\epsilon)$.

Then \\$B^\prime(m,\epsilon)=B(m,\epsilon)+\textsc{B}(m,\epsilon)$.

%%%%% and $N^\prime(\epsilon)=N(\epsilon)+\textsc{N}(\epsilon)$.

The slope as obtained from eq.(\ref{slope}) becomes,
\begin{equation}
 S(\epsilon)=\frac{ln\Lambda(\epsilon)}{ln(\frac{\epsilon}{L})}=\frac{ln\Bigg\lbrace\frac{\displaystyle\sum_{m=1}^{M(\epsilon)}m^2\frac{B^\prime(m,\epsilon)}{N^\prime(\epsilon)}}{\Big[\displaystyle\sum_{m=1}^{M(\epsilon)}m\frac{B^\prime(m,\epsilon)}{N^\prime(\epsilon)}\Big]^2}\Bigg\rbrace}{ln(\frac{\epsilon}{L})}
\label{S-gb}
\end{equation}

Further,
%%%%%%%%%%%%%%%%%%%%%%%%%%%%%%%%%%%%%%%%%
\begin{equation}
\begin{aligned}
\displaystyle\sum_{m=1}^{M(\epsilon)}m^2B^\prime(m,\epsilon)       &= \displaystyle\sum_{m=1}^{M(\epsilon)}m^2\lbrace B(m,\epsilon)+\textsc{B}(m,\epsilon) \rbrace \\
        &= \displaystyle\sum_{m=1}^{M(\epsilon)}m^2B(m,\epsilon)+\displaystyle\sum_{m=1}^{M(\epsilon)}m^2\textsc{B}(m,\epsilon) \\
        &= \lbrace 1+f(\epsilon)\rbrace \displaystyle\sum_{m=1}^{M(\epsilon)}m^2B(m,\epsilon)
\end{aligned}
\label{relat-gb}
\end{equation}
%%=====================================
 where,
 $f(\epsilon)=\frac{\displaystyle\sum_{m=1}^{M(\epsilon)}m^2\textsc{B}(m,\epsilon)}{\displaystyle\sum_{m=1}^{M(\epsilon)}m^2B(m,\epsilon)}$
 
By the similar argument we write:
%%%%%%%%%%%%%%%%%%%%%%%%%%%%%%%%%%%%%%%%
\begin{equation}
\begin{aligned}
\displaystyle\sum_{m=1}^{M(\epsilon)}mB^\prime(m,\epsilon)       &= \displaystyle\sum_{m=1}^{M(\epsilon)}m\lbrace B(m,\epsilon)+\textsc{B}(m,\epsilon)\rbrace \\
        &= \displaystyle\sum_{m=1}^{M(\epsilon)}mB(m,\epsilon)+\displaystyle\sum_{m=1}^{M(\epsilon)}m\textsc{B}(m,\epsilon) \\
        &= \lbrace 1+\textsc{f}(\epsilon))\rbrace \displaystyle\sum_{m=1}^{M(\epsilon)}mB(m,\epsilon)
\end{aligned}
\label{rel-gb1}
\end{equation}
%%=====================================
 where,
 $\textsc{f}(\epsilon)=\frac{\displaystyle\sum_{m=1}^{M(\epsilon)}m\textsc{B}(m,\epsilon)}{\displaystyle\sum_{m=1}^{M(\epsilon)}mB(m,\epsilon)}$

Substituting eqs.(\ref{relat-gb} and \ref{rel-gb1}) in eq.(\ref{S-gb}),
%%%%%%%%%%%%%%%%%%%%%%%%%%%%%%%%%%%%%%%
\begin{equation}
\begin{aligned}
S       &= \frac{ln\Bigg\lbrace\frac{N^\prime(\epsilon)\lbrace 1+f(\epsilon)\rbrace\displaystyle\sum_{m=1}^{M(\epsilon)}m^2B(m,\epsilon)}{\Big[\lbrace 1+\textsc{f}(\epsilon)\rbrace\displaystyle\sum_{m=1}^{M(\epsilon)}mB(m,\epsilon)\Big]^2}\Bigg\rbrace}{ln(\frac{\epsilon}{L})}  \\
        &= \frac{lnN^\prime(\epsilon)}{ln(\frac{\epsilon}{L})}+\frac{ln\Bigg\lbrace\frac{\displaystyle\sum_{m=1}^{M(\epsilon)}m^2B(m,\epsilon)}{\Big[\displaystyle\sum_{m=1}^{M(\epsilon)}mB(m,\epsilon)\Big]^2}\Bigg\rbrace}{ln(\frac{\epsilon}{L})}+ \frac{ln\frac{\lbrace 1+f(\epsilon)\rbrace}{\lbrace 1+\textsc{f}(\epsilon)\rbrace ^2}}{ln(\frac{\epsilon}{L})}  \\
        &= \frac{lnN^\prime(\epsilon)}{ln(\frac{\epsilon}{L})}+ D_2+\frac{ln\frac{\lbrace 1+f(\epsilon)\rbrace}{\lbrace 1+\textsc{f}(\epsilon)\rbrace ^2}}{ln(\frac{\epsilon}{L})} 
\end{aligned}
\label{rel-S}
\end{equation}
%%=====================================

From figure (\ref{glb}) it is evident that if we use gliding box counting method for lacunarity calculation,
\begin{equation}
N^\prime(\epsilon)= (L-\epsilon+1)^E=\left[(\frac{L}{\epsilon})(\epsilon-\frac{\epsilon^2}{L}+\frac{\epsilon}{L})\right]^E
\label{N}
\end{equation}

The relation given by eq.(\ref{N}) is understood better by the examination of fig.(\ref{glb}b) for a box size $\epsilon=2$ and $L=6$. On gliding the box along y-direction for a fixed value of x, we need 6-2+1 number of boxes. From fig.(\ref{glb}c), for a box size $\epsilon=3$ and $L=6$, we need 6-3+1 boxes, and so on. 

Therefore substituting eq.(\ref{N}) in eq.(\ref{rel-S}) we get,
%%%%%%%%%%%%%%%%%%%%%%%%%%%%%%%%%%%%%%%%%
\begin{equation*}
\begin{aligned}
S       &= \frac{ln\left[(\frac{L}{\epsilon})(\epsilon-\frac{\epsilon^2}{L}+\frac{\epsilon}{L})\right]^E}{ln(\frac{\epsilon}{L})}+ D_2+\frac{ln\frac{\lbrace 1+f(\epsilon)\rbrace}{\lbrace 1+\textsc{f}(\epsilon)\rbrace ^2}}{ln(\frac{\epsilon}{L})} \\
        &= \frac{ln\left[\frac{L}{\epsilon}\right]^E}{ln(\frac{\epsilon}{L})}+E\frac{ln\left[\epsilon-\frac{\epsilon^2}{L}+\frac{\epsilon}{L}\right]}{ln(\frac{\epsilon}{L})}+ D_2+\frac{ln\frac{\lbrace 1+f(\epsilon)\rbrace}{\lbrace 1+\textsc{f}(\epsilon)\rbrace ^2}}{ln(\frac{\epsilon}{L})} \\
        &= -E+E\frac{ln\left[\epsilon(1-\frac{\epsilon}{L}+\frac{1}{L})\right]}{ln(\frac{\epsilon}{L})}+ D_2+\frac{ln\frac{\lbrace 1+f(\epsilon)\rbrace}{\lbrace 1+\textsc{f}(\epsilon)\rbrace ^2}}{ln(\frac{\epsilon}{L})}
\end{aligned}
\label{S-rel1}
\end{equation*}
%%=====================================

i.e.,
\begin{equation}
S = -E+D_2+E\frac{ln\left[\epsilon(1-\frac{\epsilon}{L}+\frac{1}{L})\right]}{ln(\frac{\epsilon}{L})}+\frac{ln\frac{\lbrace 1+f(\epsilon)\rbrace}{\lbrace 1+\textsc{f}(\epsilon)\rbrace ^2}}{ln(\frac{\epsilon}{L})}
\label{S-rel3}
\end{equation}
or,
\begin{equation}
S = E(\frac{ln\left[\epsilon(1-\frac{\epsilon}{L}+\frac{1}{L})\right]}{ln(\frac{\epsilon}{L})}-1)+ D_2+\frac{ln\frac{\lbrace 1+f(\epsilon)\rbrace}{\lbrace 1+\textsc{f}(\epsilon)\rbrace ^2}}{ln(\frac{\epsilon}{L})}
\label{S-rel2}
\end{equation}
Equation(\ref{S-rel2}) is the general relation between the euclidean dimension, the generalised dimension and the slope of the lacunarity curve. For a monofractal, $D_2= D_0$, the fractal dimension. The last two terms of eq.(\ref{S-rel3}) are the additions that are necessary to the slope of the lacunarity curve as was proposed earlier by the eq.(\ref{slope2}) of Allain et.al. As these two terms contain information of the mass distribution in the factors f($\epsilon$) and $\textsc{f}$($\epsilon$), they are necessary to describe multifractal systems. With the inclusion of these terms the lacunarity curve is no longer linear for a multifractal system.

\section{Results}
\begin{figure}[!h]
\begin{center}
\includegraphics[scale=0.4]{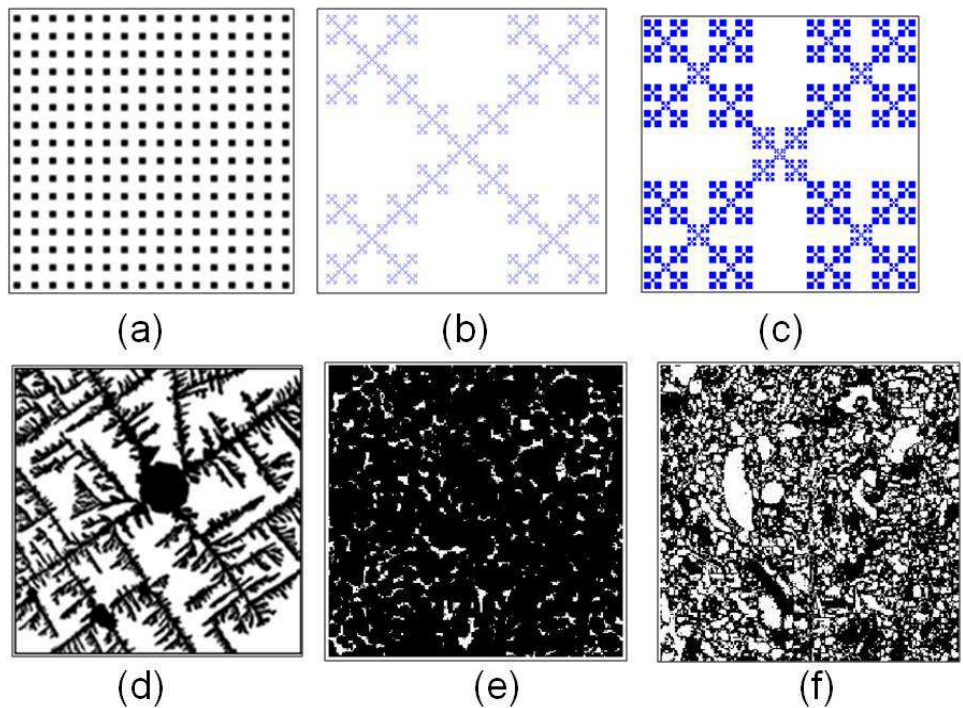} 
\caption{(a)Square grid (b)Vicsek monofractal pattern (c) Vicsek multifractal pattern (d) Aggregate of NaCl formed from solution of NaCl in  aqueous gelatin (e)Low porosity oolitic limestone from the Mondeville formation of Middle Jurassic age (Paris Basin, France) (f) High porosity reefal carbonate obtained from Majorca Islands, Spain. The white regions represent voids and the black regions, matrix. }\label{lac-case}
\end{center}
\end{figure}
\begin{figure}[!h]
\begin{center}
\includegraphics[scale=0.4]{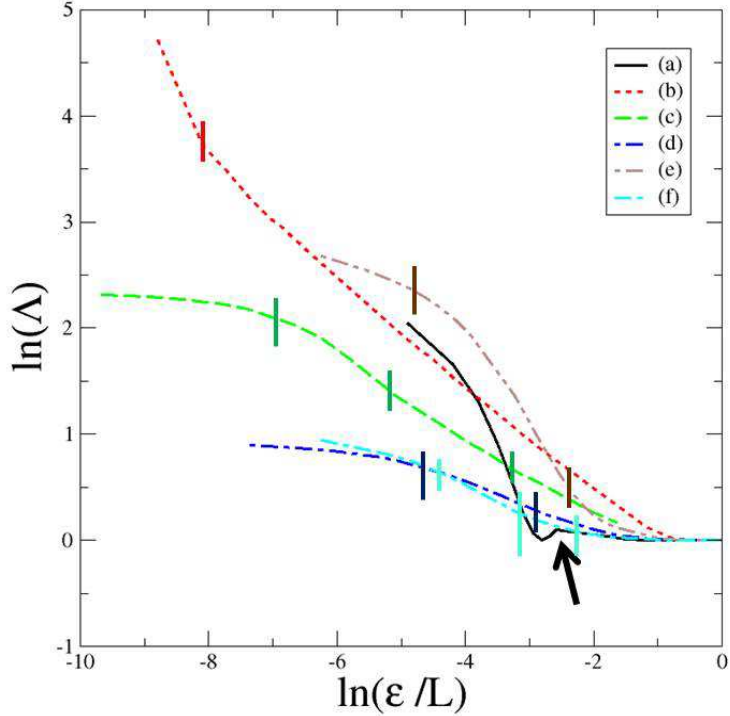} 
\caption{Lacunarity versus $\epsilon/L$ for : (a)Square grid , the arrow indicates the length scale of the periodic array (b)Vicsek monofractal (c) Vicsek multifractal (d) Aggregate of NaCl formed from solution of NaCl in  aqueous gelatin (e )Low porosity sedimentary rock (f) High porosity sedimentary rock. The breaks in slopes indicate change in length scales for multifractal systems(c,d,e,f). }\label{lac-graph}
\end{center}
\end{figure}

We present the lacunarity in six cases, fig.(\ref{lac-case}),  having different translational invariance. Three of these have been deterministically created: (a) square grid (b) 2-dimensional Vicsek pattern which is monofractal (c) 2-dimensional multifractal Vicsek pattern, while the remaining three are real systems: (d) crystal aggregation formed from an aqueous solution of NaCl and gelatin \cite{cgd2012,colsua2013} in 2-dimensions (e) sedimentary rock  of low porosity : oolitic limestone(pure calcite)from the Mondeville formation of Middle Jurassic age (Paris Basin, France) and (f) sedimentary rock of high porosity: reefal carbonate obtained from Majorca Islands, Spain, \cite{gji2015} in 3-dimensions. We have followed the gliding box algorithm outlined above to determine the lacunarity as a function of the ratio of the gliding box size to system size. The results are displayed in fig.(\ref{lac-graph}). Graph(a) represents the lacunarity of the regular distribution of points on a grid of set(a). Once the box size $\epsilon$ is greater than the grid size $\ell$, the region indicated in the figure,  n(m,$\epsilon$) becomes a constant, i.e. the variance $m_{v}^{2}(\epsilon)$ is zero. The lacunarity of a regular array is therefore $1$ for any gliding box size larger than the unit of the repeating pattern. Graph(b) represents the lacunarity of the 2-dimensional Vicsek pattern which is a monofractal. The curve is largely linear with the deviations from non-linearity occurring for $\epsilon$ less than the size of the repeating distance. For a deterministic multifractal, case(c), the lacunarity represented by graph(c) is non-linear with different slopes for different values of $\epsilon$. The breaks in slope, indicated in the figure, may be used to identify changes of scale within a multifractal system.

\begin{figure}[!h]
\begin{center}
\includegraphics[scale=0.4]{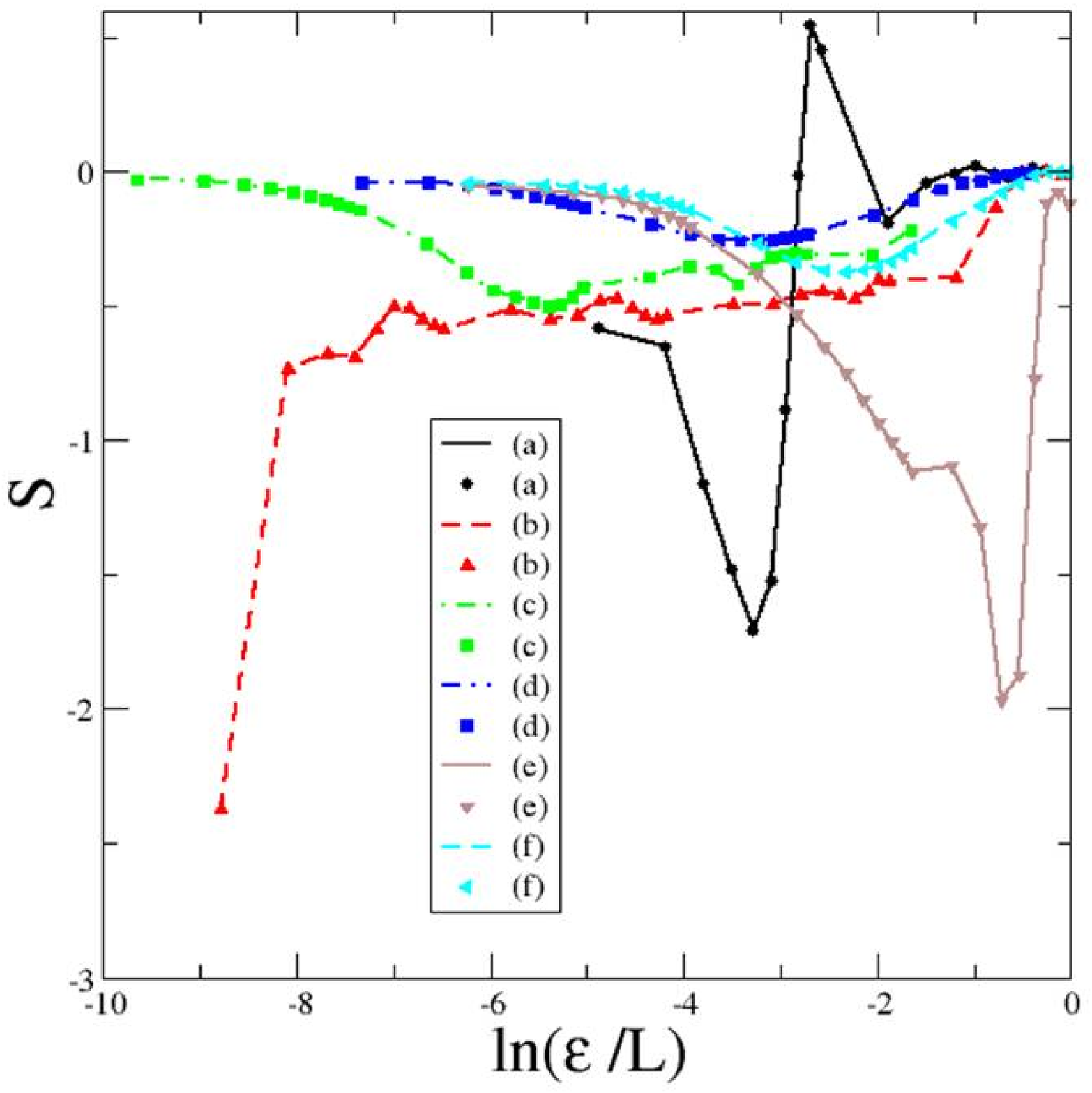} 
\caption{Comparison of lacunarity slope $S$ determined from lacunarity curves of fig.(\ref{lac-graph})(symbol), and relation of eq.(\ref{S-rel2})(lines), for the six cases:(a)Square grid (b)Vicsek monofractal (c) Vicsek multifractal(d) Aggregate of NaCl formed from solution of NaCl in  aqueous gelatin (e)Low porosity sedimentary rock (f) High porosity sedimentary rock.}\label{lac-graph-s}
\end{center}
\end{figure}
Graphs (d), (e) and (f) represent the lacunarity of real systems: graph(d) shows the lacunarity variation of the aggregation pattern (fig.(\ref{lac-case}d)) observed when a droplet of aqueous solution of NaCl and gelatin is allowed to desiccate in room conditions, 
graphs(e) and (f) represent the lacunarity of a low porous and a high porous sedimentary rock sample whose cross-sectional views are shown in figs.(\ref{lac-case}e and f) respectively. The porosity and geometry of the pore structure in sedimentary rocks is of primary importance in studies involving transport of fluid through them \cite{abhra3d, suptijh, jh2015}. The porosity of the low porosity sedimentary rock (oolitic limestone) was determined and found to be $0.074$, while the high porosity sedimentary rock (reefal carbonate)
was found to have a porosity of $0.399$ \cite{gji2015}. This high contrast in their porosity
values is evident from the panels (e) and (f) of fig.(\ref{lac-case}). The cross-sections of the rocks were obtained by suitably gray scaling binary files generated from micro-tomographs of sections of these rocks. The details of the procedure are given in \cite{gji2015, abhra3d}. In calculating the lacunarity of the sedimentary rocks, we treated the rock matrix as the `lacunae' in the system. Therefore in the implementation of both the box counting algorithm and the gliding box algorithm, we counted the number of `void' pixels in a box of size $\epsilon$. For the low porous rock, the low value of `mean void mass' and the low variance of the void distribution, together result in a higher value of lacunarity (eq.(\ref{lacu-interp})). In comparison, the sedimentary rock of higher  mean porosity, case (f), shows a lower lacunarity. For any of the two rock samples, when $\epsilon << L$, the boxes are mostly either full or empty. This results in a larger variance and therefore the lacunarity; however once the box reaches a size where the translational invariance starts to appear, lacunarity declines rapidly. For all systems when $\epsilon/L = 1$, the variance becomes zero and lacunarity becomes a constant of value $1$.

Having calculated the lacunarity of the six systems discussed above, we try to validate eq.(\ref{S-rel2}) for each. $Table I$ gives the euclidean dimension and the first (fractal dimension) and second (correlation dimension) generalised dimension where applicable. The generalised dimensions for the Vicsek multifractal was calculated following the box counting method. The relevant data for the other cases are taken from \cite{gji2015, cgd2012, Vicsek}. Using these values we calculated the slope $S$ for different values of $\epsilon/L$. Figure (\ref{lac-graph-s}) compares the value of $S$ from eq.(\ref{S-rel2}) with the slopes that are determined from the sets of fig.(\ref{lac-graph}) at the corresponding values of $\epsilon/L$. The comparison gives a very good match.

\section{Discussions and Conclusions}
Apart from the fractal dimension of monofractals and multifractals, the lacunarity measure is used to describe systems that lack translational invariance. In fact systems that have identical fractal dimension may use lacunarity to distinguish between their different textures. While the box counting method is widely used to determine the fractal dimension, the gliding box algorithm is one of the standard means to determine the lacunarity of systems. It is always interesting and necessary to know the relation if any, between these parameters that are used to describe the microstructure of systems.

In this work we first determine the lacunarity in six systems following the gliding box algorithm. The six systems have different geometry: euclidean, monofractal and multifractal; and while three of these are deterministic, three others are real systems. For the multifractal systems, the lacunarity curve is found to be non-linear. The generalised dimension of each of these systems have been determined following the box counting method and are known from previous studies. Starting from their definitions and using statistical mechanics, we then establish the relation between the euclidean dimension, the Correlation Dimension and the lacunarity for systems that lack translational invariance . In this manner we establish a bridge between the two standard methods- the box counting algorithm and the gliding box algorithm, through eq.(\ref{S-rel2}) that describes the slope of the log-log graph of lacunarity $\lambda$ versus the ratio of box size to system size, i.e. $\epsilon/L$.  

To validate our claim, we calculate the slope at different points of the log-log plot $\lambda$ versus $\epsilon/L$ and compare with the theoretical value as defined by eq.(\ref{S-rel2}) for each of the six cases we studied. The values match to within $0.01\%$, thus establishing the proposed relation. 

To conclude, we have established: (i) the lacunarity curve for multifractal systems is non-linear (ii) an exact relation between the Correlation dimension, the lacunarity and the embedding euclidean dimension of a system (iii) the proposed relation is a valid for any geometry - euclidean,  mono-fractal or multifractal.

\section{Acknowledgement} The authors are grateful to Philippe Gouze for the binary data of the micro tomographs of the sedimentary rocks. A. G. is grateful to Moutushi Duttachoudhury for the experimental photograph of desiccated sodium chloride under the microscope.

%\\
\newpage
\begin{table}
\caption{Values of the euclidean dimension and the relevant generalised dimension for the different cases. }
\begin{tabular}{|c|c|c|c|}
\hline Cases & E & D(0) & D(2) \\
\hline (a) & 2 & 2.000 & 2.000  \\
\hline (b)& 2 & 1.465 & 1.465\\
\hline (c) & 2 & 1.76833 & 1.76661\\
\hline (d) & 2 & 2.000 & 1.825 \\
\hline (e) & 3 & 2.473 & 2.396 \\
\hline (f) & 3 & 2.868 & 2.806 \\
\hline            			
\end{tabular}
\end{table}


\begin{thebibliography}{99}
\bibitem{Dutta2003}T.Dutta and S.Tarafdar, J. Geophysical Res.\textbf{108}108, NO. B2, 2062, (2003)$doi:10.1029/2001JB000523$\\
\bibitem{mandelbrot}B. Mandelbrot, The Fractal Geometry of Nature,(Freeman,
New York, 1983). \\
\bibitem{Allain1991} C. Allain, M. Cloitre,  Phys. Rev. A  \textbf{44}, 3552-3558,(1991) \\
\bibitem{Plotnick1996}R.E.Plotnick, R.H.Gardner, W.W.Hargrove, K.Prestegaard, M. Perlmutter, Phys. Rev. E  {\bf 53}, 5461-5468, (1996).\\
\bibitem{Lin1986}B. Lin and Z. R. Yang, J. Phys. A \textbf{19}, L49 ,(1986).\\
\bibitem{Gefen1984}Y. Gefen, Y. Meir, and A. Aharony, Phys. Rev. Lett. \textbf{50}, 145
,(1983). \\
\bibitem{gji2015}A.Giri, P.Gouze, S. Tarafdar, T. Dutta, Geophysical Journal International \textbf{200}, no.2P, 1106-1115, (2015)\\
\bibitem{cgd2012}A.Giri, M.D.Choudhury, T. Dutta, S. Tarafdar,  Cryst. Growth Des. \textbf{13}, 341-345, (2013).\\
\bibitem{colsua2013} T. Dutta, A. Giri, M. Dutta Choudhury and S. Tarafdar,
 Colloids
Surf., A, \textbf{432}, 127–131, 2013.\\ 
\bibitem{Vicsek}T.Vicsek, {\it Fractal Growth Phenomena, 2nd ed.}; World Scientific, Singapore, (1992) \\
\bibitem{Feder}J.Feder, {\it Fractals}; Plenum Press, New York, 1988\\
\bibitem{Diaz}M. C. Díaz , D.Giménez, A.M.Tarquis , J.M. Gascó, A. Saa  {\it Scaling Methods in Soil Physics};  Selim H. M., Pachepsky Y., Radcliffe D. E., Eds.; CRC Press:  Boca raton, Chapter 2, pp 19–33,( 2003). \\
\bibitem{Shannon1949} C.E.Shannon \& W. Weaver, \textit{The Mathematical Theory of Communication}, University of Illinois Press, Chicago, (1949)\\
\bibitem{Grassberger1983}P. Grassberger \& I. Procaccia, Physical Review Letters, \textbf{50}(5), 346–349, (1983) $DOI:10.1103/PhysRevLett.50.346$. \\
\bibitem{abhra3d}A.Giri, S.Tarafdar, P.Gouze, T.Dutta Geophysical J. Int. \textbf{192} (3), 1059-1069, 2012 $doi: 10.1093/gji/ggs084$. \\
\bibitem{suptijh}Supti Sadhukhan, Philippe Gouze, Tapati Dutta, J.Hydrology, (2012), http://dx.doi.org/10.1016/j.jhydrol.2012.05.024\\
\bibitem{jh2015}Supti Sadhukhan, Philippe Gouze, Tapati Dutta, Journal of Hydrology, \textbf{519},2101-2110, (2014)\\

\end{thebibliography}
\end{document}